\documentclass{article}
\usepackage[utf8]{inputenc}
\usepackage[pdftex]{graphicx}
\usepackage{fancyheadings}
\title{\huge \bf The normaly distributed daily returns in stock trading}

\author{ \bf Younes BEN-GHABRIT \\
\ \textit{Universit\'e du Qu\'ebec à Montr\'eal}
\\ \textit{ben-ghabrit.younes@courrier.uqam.ca}}
\begin{document}
\maketitle
\tableofcontents
\begin{abstract} In this report, we talked about a new quantitative strategy for choosing the optimal(s) stock(s) to trade. The basic notions are generally very known by the financial community. The key here is to understand 1 ) the standard score applied to a sample and 2) the correlation factor applied to different time series in real life. These notions are the core of our research. We are going to begin with the introduction section. In this part, we talked about variance, covariance, correlation factor, daily returns in stock trading and the Shapiro-Wilk test to test the normality of a time serie. Next to that, I talked about the core of my method (what do you do if you want to pick the optimal(s) stock(s) to trade). At the end of this report, I talked about a new idea if you want to analyze more than one stock at the time. All my work goes with a primary reflexion : forecasting a stock direction is a random walk and nobody can be 100 \% sure where a stock is going. All we can do, is to pretend to have a technic with a win/loss ratio greater than 51 \%.
\end{abstract}
\textbf{Keywords :} trading, strategy, returns, normality.
\section{Introduction}

First of all, the ultimate goal of this research was to correct one big error comming from the traders community : It's to trust the technical indicators like they trust there parents. They don't understand that it's just a mathematical tool that they have to master ! For exemple, if the RSI is at the 90 level, that's not automatically the sign of a reversal price action. It's just the result of some arithmetic operations. The price action is a random walk (the next price is not dependant of the historical data but he only depands of the actual price).
\\
\\
Secondly, I analyzed the daily returns of two different stocks (RIMM and AAPL). The importation and the analyze of data was with R because of the simplicity of that programmation language. I test the normality for the AAPL density function to use it for evaluating the probability of gain. Because we know that it's normally distributed, we can get a probability number going with the actual daily return of a stock (with the standard score).
\\
\\
To understand my model, you have to know some basic concepts of statistics. After this, all you have to do is to apply what you learned to a time serie and the job is done !
\subsection{Variance, covariance and correlation factor}
\ If X and Y are two random variables, then the variance of each variable is
\begin{equation}
S_x^2 = \frac{{\sum_i({x_i - \bar{x}}})^2}{n-1}
\end{equation}
\begin{equation}
S_y^2 = \frac{{\sum_i({x_i - \bar{y}}})^2}{n-1}
\end{equation}
\ If we take the square root of both sides, we get the standard deviation.

\begin{equation}  
   S_x = \sqrt{\frac{{\sum_i({x_i - \bar{x}}})^2}{n-1}}
\end{equation}
\begin{equation}  
   S_y = \sqrt{\frac{{\sum_i({y_i - \bar{y}}})^2}{n-1}}
\end{equation}
\\
\ The covariance between two jointly distributed real-valued random variables X and Y is defined as
\begin{equation}Cov(X,Y) = \frac {\sum_i ({x_i - \bar {x}})({y_i - \bar {y}})}{n-1}
\end{equation}
\ then, the correlation factor is equal to
\begin{equation} Corr(X,Y) = \frac{Cov(X,Y)}{{S_x}{S_y}} \end{equation}
\subsection{Standard score}
\ Let's define the standard score :
\ $$ Z = \frac{ X - \mu}{\sigma}$$
\ Because we are using a sample, the formula will be a little bit different
\begin{equation} Z_n = \frac {X - \bar{x}}{\frac{\sigma}{\sqrt{n}}} \end{equation}
\ $\bar{x}$ is the mean of the data for the sample and $n$ is the quantity of data we have.
\
\
\subsection{Daily returns}
\ The returns of one stock are simply expressed by the next ratio
\begin{equation} 
R_d = \frac{Close_1}{Close_0}
\end{equation}
\\
\ $Close_0$ is the yesterday's close and $Close_1$ is the today's close because our primary interest is the daily returns. We are going to modelize the price action with R \footnote{http://www.r-project.org}.
\subsection{Shapiro-Wilk test}
\ In statistics, the Shapiro–Wilk test tests the null hypothesis that a sample $x_1, ..., x_n$ came from a normally distributed population.
\ The test is : 
\begin{equation}
\ W= \frac{{(\sum_{i=1}^{n} {a_i x_i}})^2}{\sum_{i=1}^{n} ({x_i - \bar x})^2}
\end{equation}
$$(a_1,\dots,a_n) = {m^\top V^{-1} \over (m^\top V^{-1}V^{-1}m)^{1/2}}$$
\ Where $$m = (m_1,\dots,m_n)^\top\ $$
\ and $m_1, ...,m_n$ are the expected values of the order statistics of independent and identically distributed random variables sampled from the standard normal distribution, and $V$ is the covariance matrix of those order statistics.
\ We may reject the null hypothesis if W is too small (W $\leq 0.05$)
\section{Modeling}
\subsection{Modeling in R}
\ In this part, we are going to explain how can we forecast the absolute variation of two stocks \footnote{AAPL and RIMM}. Why Apple and Research in Motion ? Because they are relatively popular these days and the article will be more concrete. Let's begin with getting data from the Yahoo! Finance server. In R, we have to install the Quantmod \footnote{http://www.quantmod.com} package.
$$install.packages('quantmod')$$
\ then, we have to require it
$$ require('quantmod')$$
\ We are now able to get our data. For the purpose of this example, we select the data from the beginning of the year.
\ $$ getSymbols('AAPL', from = '2012-01-01')$$
\ $$ getSymbols('RIMM',from = '2012-01-01')$$
\ From now, the data are loaded and we need to define two variables for the daily returns.
\ $$ a = dailyReturn(AAPL)$$
\ $$ b = dailyReturn(RIMM)$$
\ To visualize the returns, we need to plot $a$ or $b$. Let's do it for $a$.
\ $$plot(a)$$

\includegraphics[scale=0.35,viewport= -180 400 50 50]{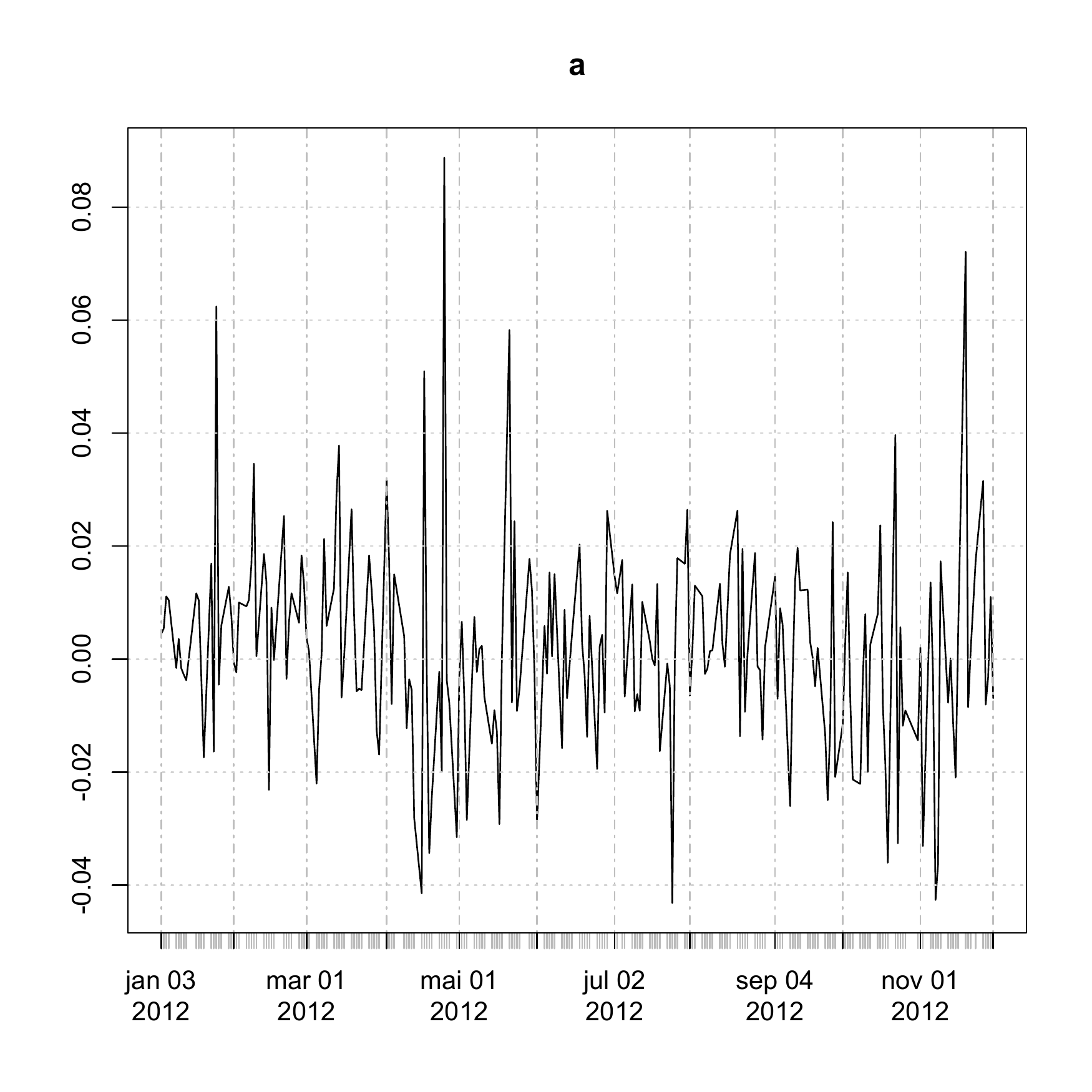}
\newpage
\ We can test the normality of that distribution with the Shapiro–Wilk test. The first thing we have to do is to plot the density of $a$ with R. $$plot(density(a))$$
\\
\\
\\
\\
\\
\\
\\
\\
\\
\\
\\
\\
\\
\\
\\
\\
\\
\\
\\
\\
\\
\\
\\
\includegraphics[scale=0.70,viewport= 10 0 50 50]{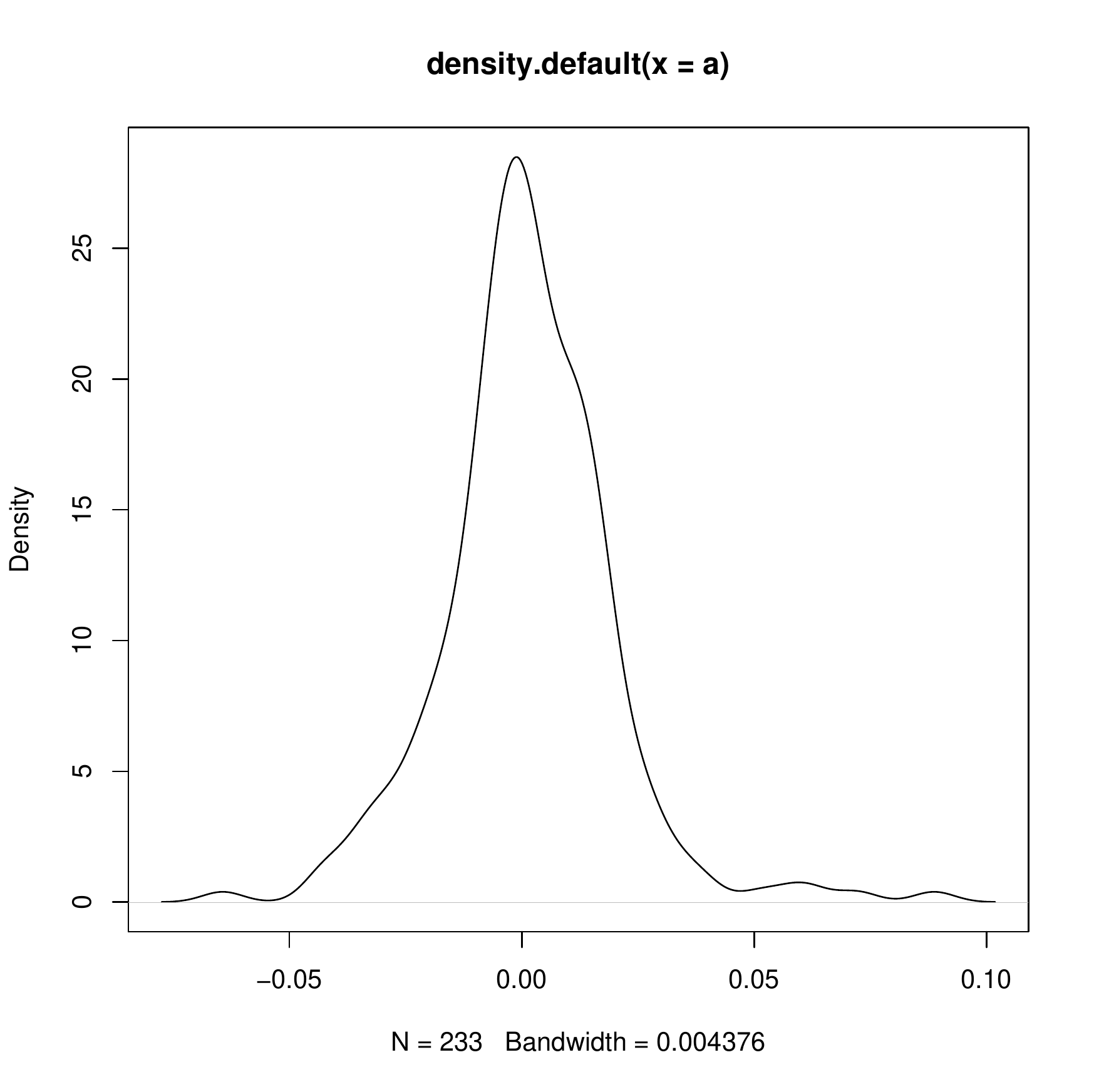}
\ To be able to test the normality of this time serie, we have to calculate $W$. But, before the computation, we have to specify the vector we want to test.
$$e=hist(a, plot=FALSE)\$breaks$$
\ $$shapiro.test(e)$$
$$W = 0.9702, p.value = 0.8924$$
\newpage
\ The QQ plot graph gives us something like that
\\
\\
\\
\\
\\
\\
\\
\\
\\
\\
\\
\\
\\
\\
\\
\\
\\
\\
\\
\\
\\
\\
\\
\\
\\
\\
\\
\\
\includegraphics[scale=0.70,viewport=  0 0 50 50]{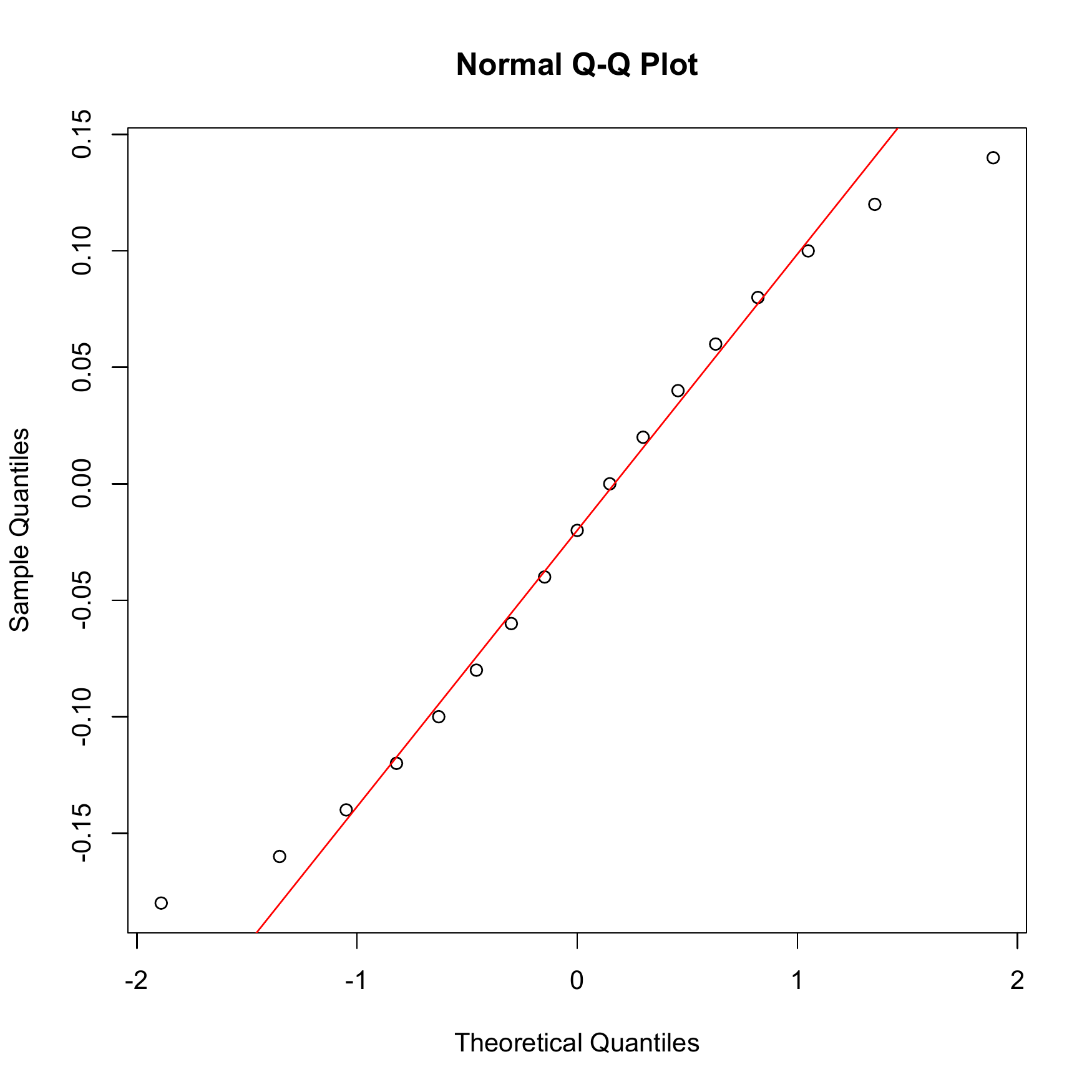}
\ We can conclude that $e$ is normally distributed.
\ With this information, we can calculate the Standard deviation with the formula $(3)$.
\\
\ The standard deviation is
$$ sd(a) = z1$$
\ The second standard deviation will be
$$ z2 = 2 * z1$$
\ Then, we add this number to the $(time;return)$ graph
$$ abline(,,z2)$$
$$ abline(,,-z2)$$
\newpage
\includegraphics[scale=0.70,viewport= 10 465 50 50]{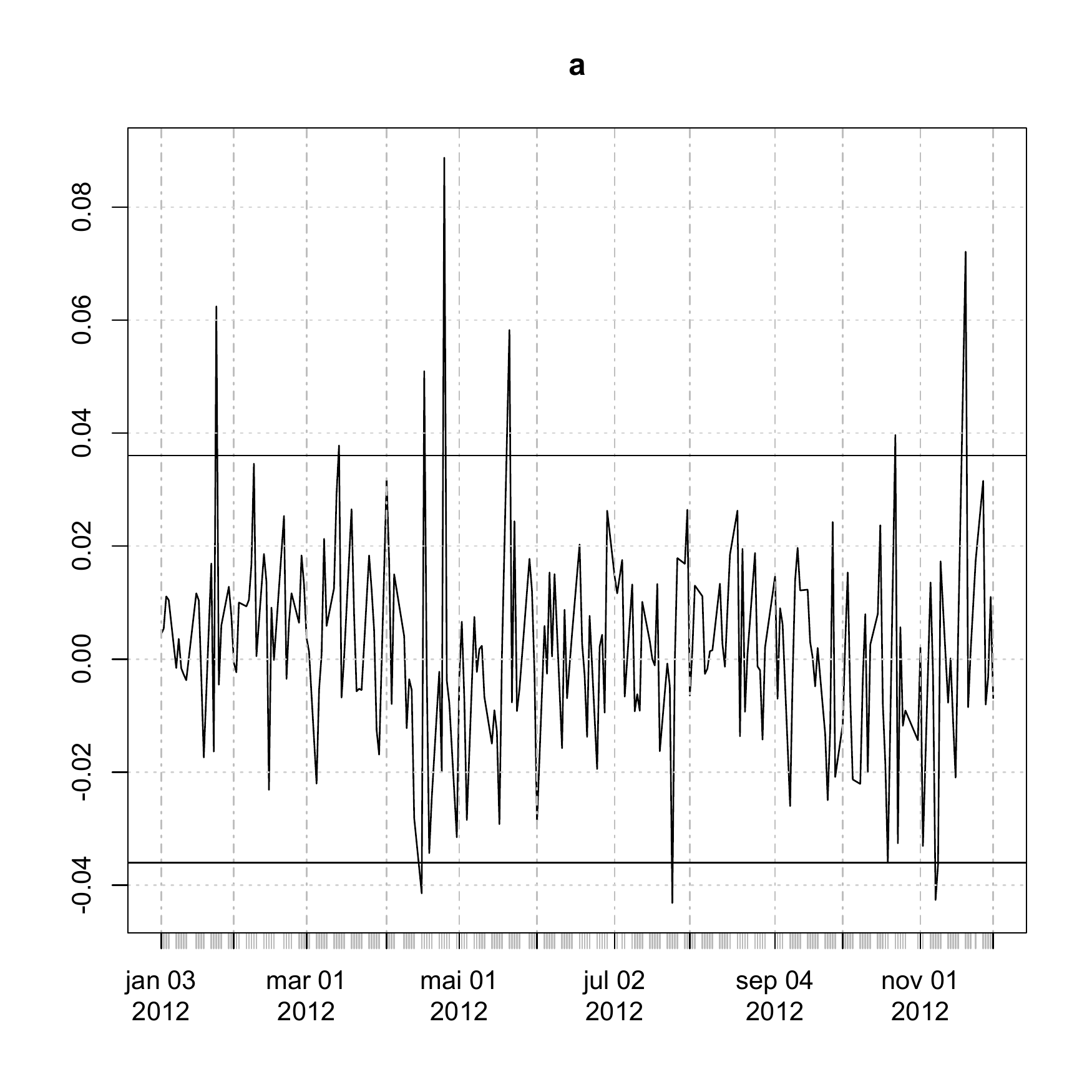}
\\
\\
\\
\\
\\
\\
\\
\\
\\
\\
\\
\\
\\
\\
\\
\\
\\
\\
\\
\\
\\
\\
\\
\\
\\
\\
\\
\\
\ With the normal distribution, we know that approximately 5 \% of the data is out of the region : $\pm 2\sigma $.
 In this case, we can trade the stock when the actual return is out of this region with a very good probility of returning in the critical zone. The daily return function can be returning in the positive or negative  zone. 78.2 \% of the tested stocks\footnote {AAPL, RIMM and YHOO from 2007-01-01 to 2012-12-01} returned in the positive zone when they was not in the critical region. The concrete application is : when a stock is getting a  high return one day above $|$$2\sigma$$|$ of the last year, there is a high probability that the next day, the return will be positive. For exemple, RIMM is 12.1 \% today and $2\sigma$ of the last year is equal to 0.045. Then we should trade RIMM tomorow and expect a return of something between 0 \% and 4.5 \%.
\\ 
\subsection{Correlation between AAPL and NASDAQ}
\ To be sure that our data of returns is correct, let's check the correlation between the returns of our stock and the returns of Nasdaq. 
\ All the steps will be with R again.
\ $$ getSymbols('^\wedge IXIC',from='2012-01-01')$$
\ $$ c=dailyReturn(IXIC) $$
\ $$ cor(a,b) $$
\\ We get a correlation of 0.6587 . The result tells us that AAPL and IXIC \footnote{Nasdaq} are correlated at 65.87 \% (a regular result because AAPL is a major stock in Nasdaq and the correlation are supposed to be greater than 50 \%).
\subsection{Momentum in the daily return graph}
\ In this section, we are going to define one differential equation that describe the momentum in the daily return graph.
\ We know that, in physics
$$ \frac{dv}{dt}=a$$
\ In finance, the momentum (acceleration in physics) is described by
\begin{equation}
\frac{dR_1}{dt} = Momentum
\end{equation}
\ We can solve this differential equation by integrating both sides
$$ \int {\frac{dR_1}{dt}} = \int{(Momentum)}$$
$$ \int {dR_1} = \int (Momentum) dt $$
$$ R_1 + c_1 = Mt + c_2 $$
\ if $c_2 - c_1 = R_0$, then the daily return of the next day is equal to
\begin{equation}
R_1 = Mt + R_0
\end{equation}
\ With this linear equation, we have to know the slope $M$, the actual date ($t$) and the today's return($R_0$).
\ We can find $M$ by doing the next operation for previous points
\begin{equation}
M =  \frac{R_2 - R_1}{t_2 - t_1}
\end{equation}
\ These informations are available in R with the function
$$ getQuote('AAPL') $$
\newpage
\section{Conclusion}
\ I will devide the conclusion section in two parts : the first part will put some conclusions into my results and the second will give you a new idea to continue this algo with fastest and better results.
\\ 
\\
\ First of all, I find two stocks who has normaly distributed daily returns and I prove it. It's a very important step (to prove it) because if you don't, you can't put any probability law into it and you will not be able to estimate any probability number. We consider that our sample data \footnote{For AAPL and RIMM} is representative. We also consider that the daily returns we got has a markovian process (A Markov process can be thought of as 'memoryless': loosely speaking, a process satisfies the Markov property if one can make predictions for the future of the process based solely on its present state just as well as one could knowing the process's full history. I.e., conditional on the present state of the system, its future and past are independent)\footnote{$http://en.wikipedia.org/wiki/Markov \_ process$}. If we concretize all that : The daily return of tomorrow is only depends on the today's daily return with the initial conditions. We also talked about the standard score : To know what stock to trade if the initial conditions are respected, you have to take the double of standard deviation of the last year and apply it to this year. For exemple, if YHOO have a daily return of $12 \% $ today and $2 \sigma$ of last year is $10 \%$, then, you have to take the trade tomorrow. The backtesting of 3 stocks in R shows us that with this algo, the probability of the daily return function for returning in a positive zone is 0.7820 (witch is a pretty good probability). If the today's daily return of YHOO is $-10 \%$, you have to do exactly the same thing that the positive case because we have a good probability of returning in positive zone. The expected daily return of this algo is between $0$ and $\sigma$. I also have to specify that this algo was tested for day-trading \footnote{$http://fr.wikipedia.org/wiki/Day-trading$}.
\\
\\
\ Finally, I have one idea to optimize the algo. The idea is to test more than 500 stocks (3 years minimum of data for each stock witch include downtrend years and uptrend years)  to be very representative of the stock market. The more stocks you analyze, the better are your predictions and the more efficient you are. I think it's a good idea to program something that gives you the ticker of stocks that follow the initial conditions and the standard score condition. With that, you can filter the stocks that do not match with your own conditions and only trade the optimal stocks.

\end{document}